# Excitation and detection of propagating spin waves at the single magnon level

Alexy D. Karenowska[*], Andrew D. Patterson, Michael J. Peterer, Einar B. Magnússon & Peter J. Leek,

Clarendon Laboratory, Department of Physics, Parks Road, Oxford OX1 3PU, UK

[*]a.karenowska@physics.ox.ac.uk

**Ferro- and ferrimagnets play host to small-signal, microwave-frequency magnetic excitations called spin waves, the quanta of which are known as magnons. Over the last decade, the field of spin-wave dynamics has contributed much to our understanding of fundamental magnetism. To date, experiments have focussed overwhelmingly on the study of room-temperature systems within classical limits. Here we demonstrate, for the first time, the excitation and detection of propagating spin waves at the single magnon level. Our results allow us to project that coupling of propagating spin-wave excitations to quantum circuits is achievable, enabling fundamental quantum-level studies of magnon systems and potentially opening doors to novel hybrid quantum measurement and information processing devices.**

Electronic spin waves are magnetic excitations found in ferro- and ferrimagnetically ordered systems. They can be thought of rather like a magnetic analogue of acoustic waves (Fig. 1a) and commonly exist at microwave (GHz) frequencies. The spin-wave quasiparticle is called the magnon; a magnon of frequency $\omega$ carries a quantum of energy $\hbar\omega$ and has a spin $\hbar$. The number of magnons populating a magnetic system is a measure of its deviation from a state of perfect magnetic order[1].

In the majority of magnetic materials, spin waves are only observed experimentally in the form of incoherent, thermal fluctuations in magnetization. However, in simple ferromagnetic metals and some ferrimagnetic garnets, they can be externally excited and propagate as coherent, dispersive signals[2,3]. The room-temperature study of travelling spin waves in magnetic waveguides — or spin-wave circuits —fashioned from such materials has long been an active area of research, widely recognized for its contribution not only to our understanding of fundamental magnetic physics, but to general wave- and quasiparticle dynamics[4-12].

Over the last 50 years, the artificially synthesized ferrimagnet yttrium iron garnet, or "YIG" ($Y_3Fe_5O_{12}$)[13,14], has been central to the advance of experimental spin-wave science. In monocrystalline form, YIG has the lowest spin-wave damping of any practical experimental material, two orders of magnitude better than the best polycrystalline magnetic metals. As a result, spin-wave propagation in waveguides made from high-quality YIG films can be observed over centimetre distances. In the experiments reported here, we study propagating spin waves in a YIG waveguide at cryogenic temperatures and demonstrate a spin-wave circuit operating at the single magnon level.

The dispersion of propagating spin-wave modes in a YIG waveguide is easily controlled via an externally applied magnetic field. Indeed, the ease with which wave parameters can be modified rapidly, substantially, and reversibly, is a feature which marks spin-wave systems out as unique against the background of those of other waves — for example, light, or acoustic excitations — where the means to make equivalent adjustments lies beyond the reach of contemporary experimental technique[6,12,14]. In a typical YIG waveguide subject to a modest bias field (< 500 mT) the most readily measurable part of the propagating spin-wave spectrum (wavenumbers $0 \leq k < 10^4$ rad cm$^{-1}$) occupies a band in the range 1 to 20 GHz.

Our experimental magnon system is an yttrium iron garnet waveguide of thickness $d = 6.5$ μm and width 2 mm grown by liquid phase epitaxy on a 0.5 mm thick gallium gadolinium garnet ($Gd_3Ga_5O_{12}$) substrate. Microwave input and output antennae (50 μm wide copper striplines) are in contact with the YIG surface (Fig. 2a). The waveguide is biased by the field from a permanent magnet, applied across its width. Such a configuration supports the propagation of surface spin waves along the length of the waveguide. The amplitude of these waves has a maximum at one surface of the waveguide from which it reduces exponentially through the thickness[2]. However, in waveguides with thicknesses of order microns, such as the one used in our experiments, the localization is typically weak, with the amplitude extending through the whole thickness. The dispersion relation for the waves is given by

$$\omega^2(k) = \omega_B(\omega_B + \omega_0) + \frac{1}{4}\omega_0^2(1 - e^{-2kd}). \quad (1)$$

Here, $k$ is the in-plane wavenumber parallel to the waveguide axis, $\omega_B = |\gamma|B$, and $\omega_0 = |\gamma|\mu_0 M_S(T)$ where $B$ is the magnitude of the bias magnetic field, $M_S(T)$ is the temperature ($T$) dependent saturation magnetization of the material, $\gamma$ is the gyromagnetic ratio, and $\mu_0$ is the permeability of free space.

**Results**

Our magnetic waveguide was mounted in a dilution refrigerator as shown in Fig. 2a. A heavily attenuated (−70 dB) microwave line was connected to the input antenna, and the output antenna was coupled to an amplified output line. Measurements were made in two configurations: the first, with a vector network analyser (VNA) between the input and output, and the second, for acquiring time-resolved data, with the input fed by a pulsed microwave source, and the output downconverted to the radio-frequency range via a quadrature IF mixer, captured using a digital to analogue converter (DAC), and envelope demodulated in software.

The right panel of Fig. 2b shows the transmission through the waveguide (S21) measured using the VNA at room temperature (black) and 10 mK (grey). The facing panel shows the corresponding theoretical dispersion curves, calculated using equation (1). In line with long-standing theory, we find the saturation magnetization of the YIG waveguide at 10 mK to be approximately 1.4 times its room-temperature value[15,16] (for our sample, $M_S(300\text{ K}) = 141.6$ kA m$^{-1}$, whilst in the low temperature limit, $M_S(10\text{ mK}) = 197.4$ kA m$^{-1}$). All else being equal, one would therefore expect the spectrum to occupy a higher band of frequencies cold than at room temperature (equation (1)). However, the field from the permanent magnet is found to be slightly lower at 10 mK than at room temperature, resulting in the net ∼ 250 MHz downward shift observed in the data. The two datasets are scaled to compensate for the temperature-dependent properties of the microwave components in the signal lines. The low-frequency limit on transmission is the ferromagnetic resonance frequency (the $k = 0$ magnon mode) at which all spins in the waveguide precess in phase. The gradual reduction in signal as the frequency increases is chiefly as a result of the fact that, due to their finite width, the efficiency of the antennae reduces as $k$ increases. It can be seen that across the majority of the spectrum, the loss across the waveguide when cold is approximately twice the room-temperature value. One might expect the opposite to be true, since the probabilities of magnon-phonon and ionic impurity scattering, and multi-magnon processes are reduced at very low temperatures[17,18]. However, the gallium gadolinium garnet substrate upon which the YIG film is grown is known to be paramagnetic at low temperatures[19-20]. The influence of this paramagnetism on the YIG system has long been recognized as a significant barrier to performing experiments on waveguides at temperatures in the Kelvin range[19-20], and it is plausible that its effects though — given the quality of our data — obviously diminished, extend down to millikelvin.

Having established and investigated continuous spin-wave transmission through the waveguide, we went on to undertake a series of time-resolved measurements to directly probe the magnon dispersion. A 25 ns wide microwave pulse was applied to the input antenna (P1, Fig. 2a), with a centre frequency swept across the full spin-wave band in steps of 1 MHz, and the envelope of the spin-wave output (P2) measured via downconversion; first to an intermediate frequency of 62.5 MHz using a quadrature (I/Q) mixer, and then to DC in software. Figure 3 shows data acquired in this way at room temperature (3a) and 10 mK (3b). The position in time of the microwave stimulus pulse, which arrives at the input to the waveguide after a delay of 34 ns from the beginning of the experiment ($t = 0$), is indicated by the translucent overlay. For ease of visual comparison, the two plots are separately normalized; the difference between absolute signal amplitudes is as predicted by the VNA data of Fig. 2b. The frequency-dependent delay is well modelled by equation (1) (superposed curves); it can be seen that the substantial increase in the saturation magnetization of the waveguide upon cooling leads to an appreciable increase in the slope of the dispersion curve (i.e the velocity of the magnons). The lower panels of Figs 2a,b show individual time traces at the frequencies indicated by the dashed white lines in the upper plots (7,100 MHz in a, 6,750 MHz in b). Here, the vertical axes show the raw downconverted signal voltages.

We can estimate the number of magnons injected into the waveguide per microwave input pulse in the experiments of Fig. 3 by dividing the total energy of the electromagnetic pulse arriving at the waveguide input by the energy of a magnon at the appropriate centre frequency. If we assume that the coupling between the electromagnetic field from the input antenna and our magnetic system is perfect, we arrive at an upper bound of $N_{\max} \approx 2{,}900$ magnons per pulse at the power level employed in these measurements. Having characterized the system at this signal level, we went on to repeat the cold measurement at progressively lower input pulse powers. In Fig. 4 we show part of this dataset illustrating the evolution of the signals as the input power is reduced from the value in Fig. 3 to the single magnon level. Time traces at each frequency are averaged over 525,000 repetitions. It should be noted that these data were collected under slightly different magnetic bias conditions, hence the small difference in the band of frequencies occupied by the magnons relative to Fig. 3.

Experimentally measured output signal powers for the whole of the measurement set from which the data of Fig. 4 are extracted are summarized in the log-log plot of Fig. 5. The powers indicated on the left-hand axis are those measured at the output of our data-acquisition system after amplification, these are calculated by averaging the square of the raw signal voltages recorded during the pulse and dividing by the system's input impedance (50 Ω). On the horizontal axis, we specify the output power from the microwave signal generator driving the experiment (i.e. the power of the 25 ns microwave pulse), prior to attenuation through the input signal line. The overall effective attenuation of the input signal line, including the effects of partial reflection of the signal from the imperfectly matched waveguide input antenna is -73 dB. The maximum number of magnons per pulse at the waveguide input can be read off the top horizontal axis. Our data is an excellent fit to a straight line of unity slope, shown in red on the plot.

**Discussion**

A simple quasi-classical model of the magnetic structure of the waveguide used in our experiments can be used to quantify the size of the magnetic fields associated with the small spin-wave signals being measured. The presence of magnons in a magnetic material is heralded by a change in its static magnetization $\Delta M_z = M_S(1 - \cos\theta)$ where $\theta$ is the notional precessional angle of the spins about the magnetization direction (Fig. 1b). An AC magnetization perturbation (a vector rotating at the frequency $\omega$ of the magnon) $\Delta M_{AC} = M_S \sin\theta$ exists in the plane perpendicular to $\Delta M_z$ and it is this that is directly measured in our experiments. For small $\theta$, $\Delta M_z \simeq M_S \theta^2/2$ and $\Delta M_{AC} \simeq M_S \theta = \sqrt{2\Delta M_z M_S}$. The change in static magnetization $\Delta M_z$ brought about by a single magnon of frequency $\omega$ within a volume $V$ is related to the magnon energy by $\hbar\omega = (\Delta M_z \cdot B)V$. The foregoing relations enable us to estimate that, under the conditions of our experiment, a single magnon of frequency 6,700 MHz gives rise to a $\Delta M_{z_{n=1}} \simeq 1.4$ pA m$^{-1}$ which is equivalent to a field $\mu_0 \Delta M_{z_{n=1}} \simeq 1.8$ aT. The corresponding AC field $\mu_0 \Delta M_{AC_{n=1}} \simeq 0.9$ nT. In our lowest power measurements, we inject a maximum of 0.3 magnons into the waveguide per pulse, and this signal is then attenuated by at least 4 dB as it propagates across the waveguide. This, the smallest signal we are able measure at the output antenna, corresponds to a static magnetization change $\mu_0 \Delta M_{z_{n=0.3}} \simeq 0.3$ aT, and a dynamic signal of $\mu_0 \Delta M_{AC_{n=0.3}} \simeq 0.2$ nT. On the basis of these calculations, it is natural to consider the prospects for a single-shot measurement of a single propagating magnon. In our system, we are operating with a receiving antenna which is broadband on the frequency-scale of the magnon system; its quality factor, $Q$, is small. Since the signal-to-noise ratio of the antenna measurement system scales as $\sqrt{Q}$, we can predict from our results that a single-shot measurement could be achieved with a receiving antenna having a $Q$ of $10^4$. Such a $Q$ is reachable using a superconducting niobium resonator[21], even in the presence of an in-plane magnetic field of order 100 mT.

Alternatively, the change in current across a superconducting quantum interference device (SQUID)[22] fabricated on the surface of a YIG waveguide could be used to detect the magnetic field of a magnon passing beneath it.

The fact that we can observe single-magnon level propagating spin waves at millikelvin temperatures further points to the possibility of coupling such signals to superconducting quantum circuits — structures in which microwave-frequency electromagnetic signals are manipulated at the quantum level for applications such as quantum computing and metrology[23]. A commonly employed superconducting quantum circuit architecture involves Josephson junction based quantum bits (qubits) embedded in high-quality microwave resonators to form a circuit analogue of cavity quantum electrodynamics (QED)[24,25]. Latterly, progress has been made in coupling superconducting quantum circuits to other physical systems[26] such as spin ensembles[21,27], quantum dots[28,29], and mechanical resonators[30,31]. As well as providing a means to transfer quantum information between different media, such devices are a powerful tool in fundamental physical studies. Recently, a bulk (i.e. non-propagating, $k = 0$) spin-wave mode of a YIG sphere has been coupled to a 3D microwave cavity[32], and also to a superconducting qubit[33].

The existing hybrid systems discussed above all involve static, or resonant quantum systems. Coupling to a propagating excitation, such as the spin wave which forms the focus of this present work, potentially supports a wider range of functionalities including, for example, sophisticated spatial and temporal addressing. Quantum circuits incorporating propagating spin waves could not only offer new ways to tackle fundamental questions in low-energy magnon dynamics but, by virtue of the unusual and highly tunable dispersion of the excitations, open up the possibility of creating a quantum signaling environment that is uniquely controllable. It is interesting to compare the propagating spin wave with the surface acoustic wave, which has recently been measured at the single-phonon level on piezoelectric GaAs using a single electron transistor[34] and observed in interaction with a superconducting qubit[35]. The wavelengths and velocities of both surface acoustic and spin waves are many orders of magnitude smaller than those of electromagnetic waves. This feature, which already underpins their use in diverse classical analogue signal processing devices[36,37] will also make it possible to observe and manipulate on-chip quantum signals in novel and potentially technologically useful ways.

In summary, we have performed a detailed low-temperature study of dispersive magnon modes in the ferrimagnet yttrium iron garnet and report three main results: Firstly, confirmation that the ~0 K region of the $M(T)$ curve of this important experimental material is consistent with long-standing theory. Secondly, demonstration that though — as has been known for some time — magnon propagation in YIG waveguides at temperatures in the Kelvin range is inhibited by paramagnetism in the gallium gadolinium garnet substrate upon which films of the material are conventionally grown[19-20], it prevents no barrier to quantum experiments at millikelvin temperatures —hitherto almost entirely uncharted experimental territory. Thirdly, the excitation and detection of propagating spin waves at the single magnon level. On the strength of our investigations, we observe that the experimental tools now exist to allow us to fabricate microwave quantum circuits incorporating dispersive magnon systems. With this possibility comes the promise both of unpicking quantum aspects of magnon physics with unprecedented clarity, and of investigating how this physics — in particular the magnon's highly manipulable dispersion[12,14], its accessible non-linearity[14], and its ability to couple to optical[38-40] and electron-based spintronic systems[11,41-47] — could play a role in future microwave quantum technologies.

**Acknowledgements**

The technical assistance of Bob Watkins is gratefully acknowledged, as are fruitful discussions with Prof. John Gregg at the University of Oxford, and the group of Prof. Daniel Esteve at CEA, Saclay. This work was supported by the Engineering and Physical Sciences Research Council under grants EP/K032690/1, EP/J001821/1, and EP/J013501/1. ADK also is appreciative of support from Magdalen College, Oxford.

**Author contributions**

ADK conceived of, designed, and performed the experiments, analyzed the experimental data, and wrote the paper. ADP, MJP, and EBM contributed the development of the low-temperature measurement setup under the supervision of PJL. PJL also assisted with the experiments, and with the writing of the paper.

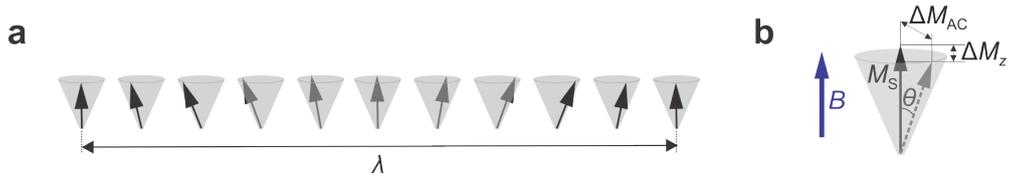

**Figure 1| A quasi-classical picture of spin-wave dynamics.** (**a**) A spin wave is an excitation of the spin-lattice of a magnetically ordered material which can be described as a harmonic variation (wavelength $\lambda$) in the phase of precession of adjacent spins about the local magnetization direction. A single magnon corresponds to a $\pi$ flip (i.e. a 180 degree rotation) of a single spin. (**b**) The change in static magnetization brought about by the presence of magnons $\Delta M_z = M_S(1 - \cos\theta)$ where $\theta$ is the notional precessional angle of the spins about the magnetization direction (parallel to the applied bias magnetic field $B$ in our experiments) and $M_S$ is the saturation magnetization. An AC magnetization perturbation (a vector rotating at the frequency $\omega$ of the magnon) $\Delta M_{AC} = M_S \sin\theta$ exists in the plane perpendicular to $\Delta M_z$. For small $\theta$, $\Delta M_z \simeq M_S \theta^2/2$ and $\Delta M_{AC} = M_S \theta$.

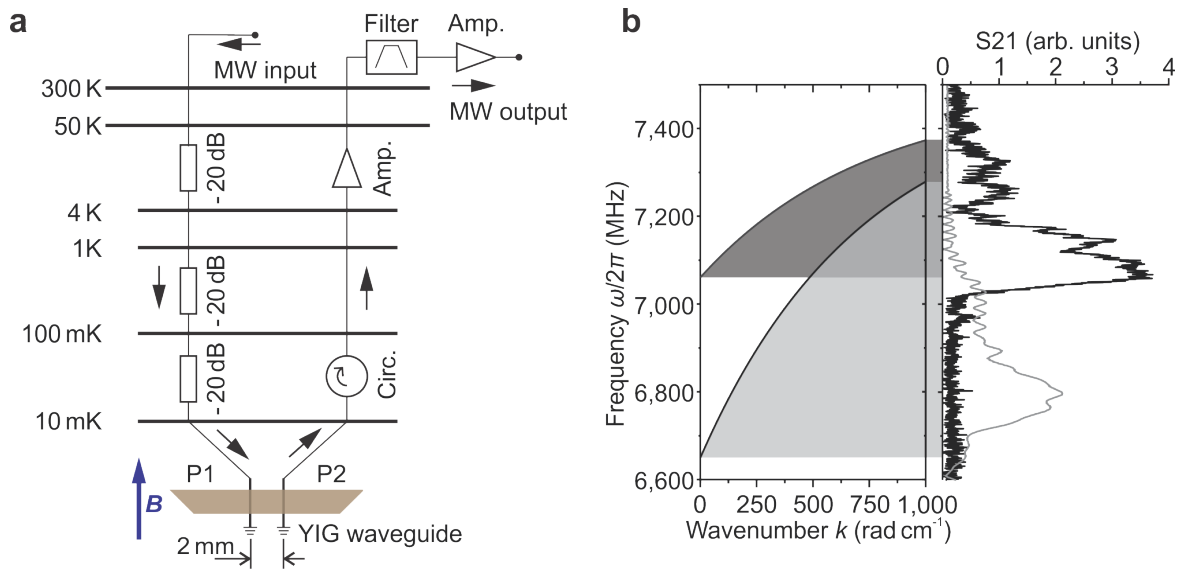

**Figure 2 | Measurement configuration and sample characterization.** (**a**) The yttrium iron garnet waveguide used in our experiments has a thickness of 6.5 μm, width 2 mm, and length 2 mm, it is biased across its width by the field from a permanent magnet. Inductive spin-wave input and output antennae are formed from 50 μm wide copper striplines. The sample assembly is mounted in a dilution refrigerator with its input (port P1) connected to a heavily attenuated coaxial microwave line (total attenuation 70 dB; 60 dB from attenuators, 10 dB from the coax). The output signal (port P2) is collected via an amplified output line. Measurements were made in two configurations: the first, with a vector network analyser connected between the input and output, and the second, for recovering time-resolved data, with the input connected to a pulsed microwave source. (**b**) The right panel shows the measured transmission through the waveguide (the s-parameter S21) at room temperature (black) and 10 mK (grey) and the left panel, the corresponding dispersion curves calculated using equation (1).

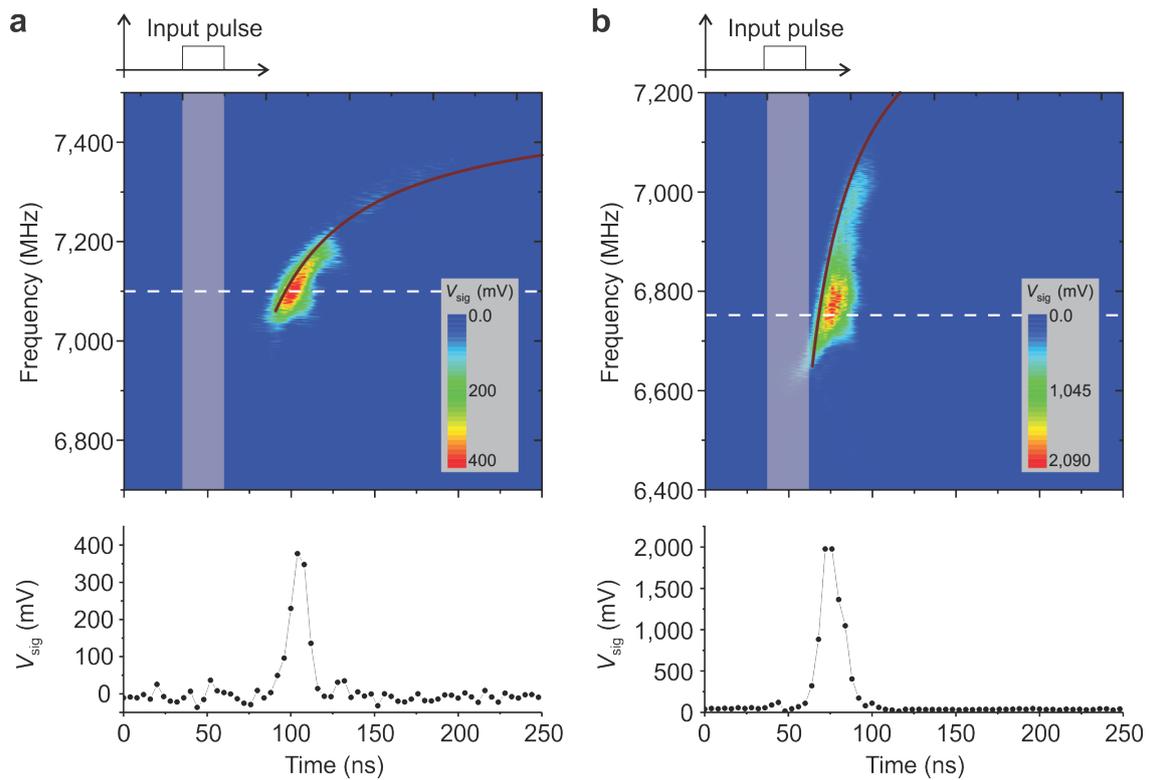

**Figure 3| A comparison of waveguide dispersion at room temperature and 10 mK.** Data displayed in the top panels were collected by applying a 25 ns pulse of varying centre frequency to the input of the waveguide (P1), and capturing the output at room temperature (**a**) and 10 mK (**b**). The colour scales reflect the downconverted signal amplitudes in volts (see scale bars) and each of the plots is separately normalized. Relative amplitudes are as predicted by the transmission data shown in Fig. 2b. The position in time of the exciting pulse is indicated by the grey translucent overlay. The superposed curves show the theoretical frequency-dependent delay calculated using equation (1). The time resolution of the measurements is 4 ns and the frequency resolution is 1 MHz. The lower panels show individual time traces at the frequencies indicated by the dashed white lines in the upper plots (7,100 MHz in (**a**), 6,750 MHz in (**b**)).

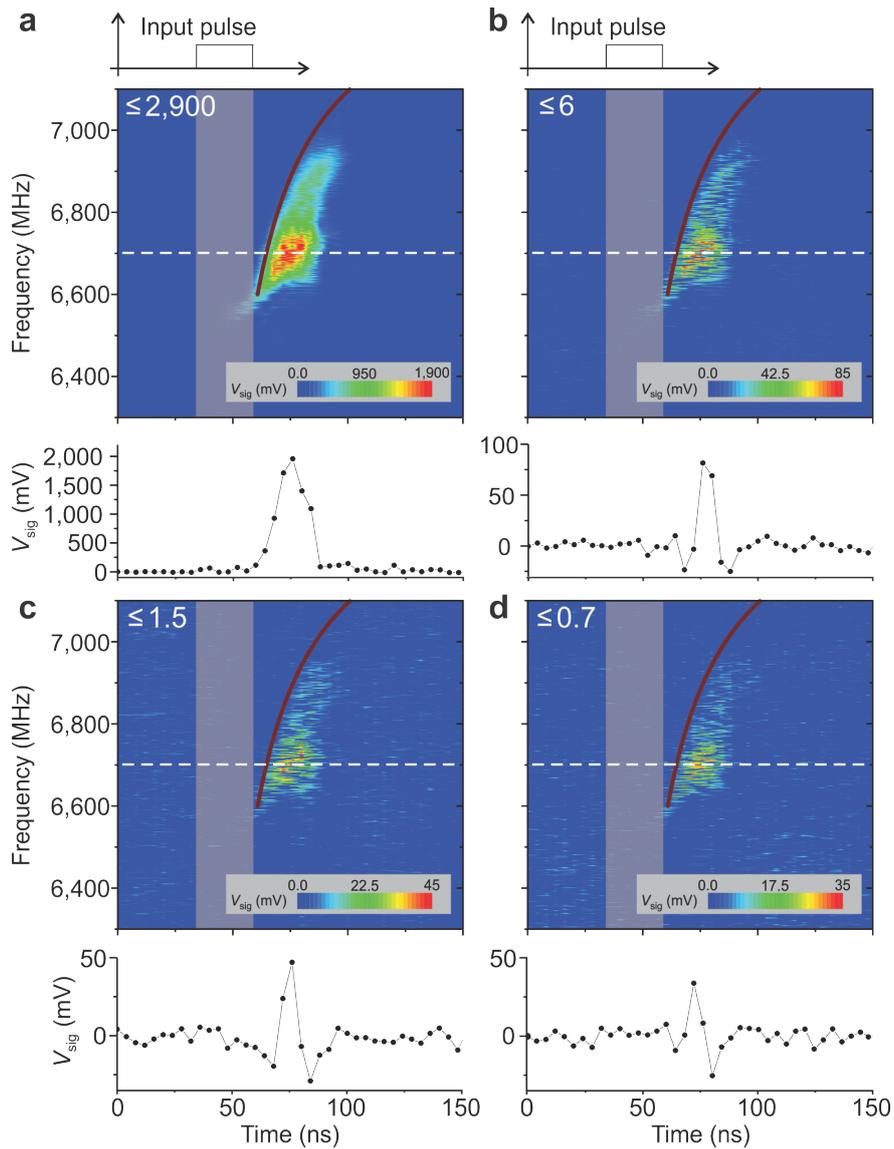

**Figure 4 | Time-resolved measurement of spin-wave propagation at the single magnon level.** Data collected in the same fashion as those of Fig. 3 as the power of the input pulse is reduced to the single magnon level. As in Fig. 3, the colour scales reflect the downconverted signal amplitudes in volts, and each plot is separately normalized. The maximum number of magnons at the waveguide input, calculated at 6,700 MHz, is indicated in the top left-hand corner of each plot. The superposed curves show the theoretical frequency-dependent delay calculated using equation (1).

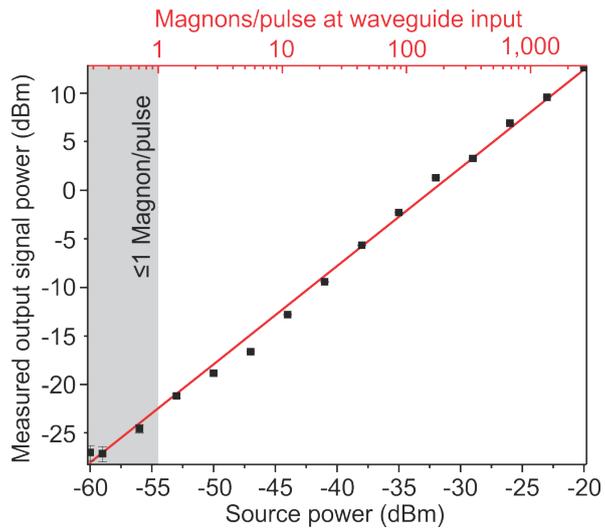

**Figure 5 | Summary of experimentally measured signal powers.** Plotted are the output signal powers in our time-resolved experiments measured at 6,700 MHz after significant amplification (black squares). The bottom horizontal axis indicates the output power of the microwave source used in each experiment (i.e. the power of the 25 ns driving pulse) prior to attenuation of through the input signal line. The data are well described by a straight line of slope unity (red). The maximum number of magnons per pulse at the waveguide input can be read off the top horizontal axis.